\documentclass[aps,prd,reprint,nofootinbib]{revtex4-2}

\usepackage{xspace}
\usepackage{amsmath}
\usepackage{graphicx}
\usepackage{csvsimple}
\usepackage{booktabs}
\usepackage{mathtools}
\usepackage{nicematrix}
\usepackage{orcidlink}

\graphicspath{{./figures/}}

\mathtoolsset{showonlyrefs}

\newcommand{\rs}{\\[0.2em]}

\newcommand{\score}{\texttt{SCoRe}\xspace} 

\newcommand{\var}{\mathrm{Var}\xspace} 
\newcommand{\Lk}{\ensuremath{\mathcal{L}}\xspace} 
\newcommand{\Po}{\ensuremath{\mathcal{P}}\xspace} 
\newcommand{\prior}{\ensuremath{\pi}\xspace} 
\newcommand{\varOfD}{\ensuremath{K_{D}}} 
\newcommand{\effL}{\ensuremath{\tilde{\Lambda}}\xspace} 
\newcommand{\bdev}{\ensuremath{\mathcal{B}^{\rm dev}_{\rm nodev}}\xspace} 

\usepackage[acronym]{glossaries}

\newacronym{snr}{SNR}{signal-to-noise ratio}
\newacronym[longplural={Effective field theories}]{eft}{EFT}{Effective field theory}
\newacronym{crisp}{CRISP}{Cross-correlated Residual Power}
\newacronym{mle}{MLE}{Maximum Likelihood Estimator}
\newacronym{gw}{GW}{Gravitational Waves}
\newacronym{gwtc3}{GWTC-3}{the Third Gravitational-Wave Transient Catalog}
\newacronym{gr}{GR}{General Relativity}
\newacronym{lvk}{LVK}{LIGO-Virgo-KAGRA}
\newacronym{ci}{CI}{creditble interval}

\newcommand{\mthreshold}{\ensuremath{m_\text{threshold}}\xspace}

\begin{document}

\title{Agnostically decoding gravitational wave model deficiencies in GWTC-3}

\date{\today}

\author{Guillaume Dideron \orcidlink{0000-0002-5222-7974}}\email{gdideron@perimeterinstitute.ca}
\affiliation{Perimeter Institute for Theoretical Physics, 31 Caroline Street North, Waterloo, ON N2L 2Y5, Canada}

\author{Suvodip Mukherjee \orcidlink{0000-0002-3373-5236}}\email{suvodip@tifr.res.in}
\affiliation{Department of Astronomy \& Astrophysics, Tata Institute of Fundamental Research, 1, Homi Bhabha Road, Colaba, Mumbai 400005, India}

\author{Luis Lehner \orcidlink{0000-0001-9682-3383}}\email{llehner@perimeterinstitute.ca}
\affiliation{Perimeter Institute for Theoretical Physics, 31 Caroline Street North, Waterloo, ON N2L 2Y5, Canada}

\begin{abstract}
\gls{gw} data bring an exceptional avenue to test the underlying models of coalescing compact objects. In the regime of strong gravity and high curvature, they allow the exploration of minute deviations from the best-fit models, which are difficult to uncover with other observational modalities.
These deviations can stem from departures from \gls{gr} or unaccounted astrophysical effects. They may not be explainable within the current
description of \gls{gw} strain data, or may simply be difficult to model.
However, they are expected to be correlated between detectors and across the 
population of observed events.
The recently developed analysis pipeline \score  leverages these properties
by focusing on the correlated power between detectors and combining results from 
multiple events.
In this paper, we apply the framework on \gls{gwtc3} to search
for source-dependent deviations.
In particular, we explore whether there is evidence for a mass-scale in the observed events, which can act like a line of demarcation in their physical properties by exhibiting a deviation that is different
above and below this mass-scale. This mass scale dependency naturally arises in gravitational theories described through effective field theories, due to environmental effects or in scenarios involving exotic compact objects, where the \gls{gw} signature can differ from the standard binary black holes in \gls{gr}. Using the 30 highest Signal-to-Noise Ratio events in the catalog, we find Bayes factors ranging from 0.16--0.5 (depending on where the threshold mass is set), thus disfavoring the hypothesis of existence of any mass-scale between $\sim 2.5$ M$_\odot$ and $60$ M$_\odot$. We also compute the distribution of excess cross-correlated power across events and find a Bayes factor of $0.07$, which agrees with expected noise statistics.

\end{abstract}

\maketitle

\glsresetall

\section{Introduction}%
\label{sec:Introduction}

\gls{gw} astronomy furnishes outstanding opportunities to learn about
our universe~\cite{2021NatRP...3..344B}. The depth to which we can answer fundamental questions is tied to our abilities to scrutinize observations
and infer relevant lessons. To date, impressive insights have been obtained (by research from within and outside the \gls{lvk} collaboration), for instance on rates and distribution of compact
binaries (e.g.~\cite{theligoscientificcollaborationGWTC40UpdatingGravitationalWave2025,callister2024observedgravitationalwavepopulations}), the consistency of General Relativity to
describe the inspiral, merger and ringdown (e.g.~\cite{LIGOScientific:2021sio,Yunes:2013dva}) and
the connection with short gamma ray bursts (e.g.~\cite{LIGOScientific:2017zic,Sarin_2022,desantis2026constrainingbinaryneutronstar}), to
name just a few.\footnote{The flip side of the excitement this era has brought is that it has become impossible to cite all
exciting developments and results in a broad context.} As the field transitions to
increasingly higher precision and with it a plethora of
events~\cite{abacGWTC40IntroductionVersion2025}, exciting opportunities become possible to explore from astrophysics to fundamental physics. On the other hand, the challenge of dealing with unmodelled contributions present in the \gls{gw} signal also becomes more pressing. The search for unmodelled signals (or features in the signals) can unearth several astrophysical and fundamental physics aspects, ranging from the nature of compact object sources and their environment to testing GR. Moreover, even when incorporating all the relevant physics governing the waveforms, deviations of the \gls{gw} data with respect to the best-fit model can still arise due to waveform systematics
(see, e.g.~\cite{LIGOScientific:2016ebw,Moore:2021eok,Yelikar:2024wzm,Gupta:2024gun,Bini:2026kwz,Mezzasoma:2026wme}).

Standard \gls{gw} studies that 
focus on deviations from \gls{gr}
(e.g.~\cite{theligoscientificcollaborationGWTC40TestsGeneral2026a,theligoscientificcollaborationGWTC40TestsGeneral2026, theligoscientificcollaborationGWTC40TestsGeneral2026b, LIGOScientific:2021sio,chiaPursuitLoveNumbers2024,Yunes:2013dva,universe7120497}) include: constraining specific
parametrised strain deviations (e.g. Post-Newtonian
coefficients~\cite{LIGOScientific:2021sio,agathosTIGERDataAnalysis2014a,Yunes:2009ke,nairFundamentalPhysicsImplications2019,yunesTheoreticalPhysicsImplications2016,mehtaTestsGeneralRelativity2023,royImprovedParametrizedTest2026}); looking for
specific physical signatures, such as modified dispersion
relations~\cite{kosteleckyTestingLocalLorentz2016,Mastrogiovanni:2020gua,okounkovaConstrainingGravitationalWave2022,shaoCombinedSearchAnisotropic2020,mirshekariConstrainingLorentzviolatingModified2012a},
spin-induced quadrupole
moments~\cite{ryanGravitationalWavesInspiral1995,poissonGravitationalWavesInspiraling1998,laarakkersQuadrupoleMomentsRotating1999,krishnenduTestingBinaryBlack2017},
or non-tensorial polarizations~\cite{wongNullstreambasedBayesianUnmodeled2021}; and
statistically significant coherent residual signal after subtracting the
best-fit waveforms, among others.
When possible, hierarchical
schemes~\cite{isiHierarchicalTestGeneral2019,zhongMultidimensionalHierarchicalTests2024} are used to combine results from multiple
events. 
Indeed, just as combining events enables
constraining frontier astrophysics, such as the mass gaps and formation channels, 
is also a powerful method to explore unmodelled physical effects that are either too
minute in individual events or only manifest as a population effect.

Motivated by population level inference of physical unmodelled effects from noisy \gls{gw} data, we developed a new optimal framework \score in previous works~\cite{dideronNewFrameworkStudy2023,dideronDetectingUnmodeledSourcedependent2025} and demonstrated its capability in measuring minute signal from the data accessible from the \gls{lvk}~\cite{KAGRA:2013rdx, LIGOScientific:2014pky, VIRGO:2014yos, KAGRA:2020tym} and third generation detectors, such as Cosmic Explorer\cite{Reitze:2019iox} and Einstein Telescope\cite{2010CQGra..27h4007P}, using simulated \gls{gw} data.
The \score framework aims to constrain deviations from the waveform and noise models
used in \gls{gw} parameter inference, assuming only that any residual power due to
the deviation is coherent between detectors. This allows making minimal assumptions on 
the strain model, a particularly important feature to enable combining a population of events
without relying on a precise model for the underlying strain deviation. In
turn, by focusing the search on population effects, we can leverage them to
differentiate between sources of deviations that may be degenerate at the
strain level. The details of this framework have
been presented in~\cite{dideronNewFrameworkStudy2023,dideronDetectingUnmodeledSourcedependent2025}. 
In this work, we employ \score for the first-time on  \gls{gw} data from
\gls{gwtc3}~\cite{ligoscientificcollaborationGWTC3CompactBinary2023} events and, to illustrate its application, we explore whether there is any signature of mass-scale in the \gls{gw} data that sets a line of demarcation between the properties of compact objects at the near field strong gravity regime. Such a mass-scale can be related to potential beyond \gls{gr} effects or the type of objects involved in the binary (such as exotic compact objects). We search for this mass-scale using both a modelled and an unmodelled approach.

We first provide a description of our formalism in Sec~\ref{sec:Method}. We then report our choice of events, and describe two population models that each resulted in their own analysis.
In the first model, we separate the events in two populations, delimited by a mass-scale, and look for differences in how excess cross-correlated power is distributed in both populations.  We find that, regardless of the mass-scale value, the two populations exhibit deviations consistent with noise. Moreover, we find that the largest support is for the two populations being identical.
In the second model, we reproduce the analysis
in~\cite{dideronDetectingUnmodeledSourcedependent2025}, and look for a
\gls{eft}-motivated deviation that scales as a power law in the total mass of the
binary. We find that the cross-correlated power is not able to place
strong enough bounds on strain-level parameters to allow inference of the power law index.

\section{The \score formalism}
\label{sec:Method}

\subsection{Cross-correlated Residual Power}%
\label{sub:Cross-correlated Residual Power}
 The \score framework \cite{dideronDetectingUnmodeledSourcedependent2025,dideronNewFrameworkStudy2023} enables the search for signatures of unmodelled physical effects in \gls{gw} data arising from beyond\gls{gr} effects, unmodeled astrophysics, or waveform systematics by analyzing the cross-correlated residuals between different detectors after subtracting best-fit signals. It identifies coherent, structured features in these residuals that could indicate unmodeled physics, as opposed to random detector noise. In essence, it provides a systematic, Bayesian way to search for new physics hidden in the small, correlated discrepancies left behind after standard \gls{gw} analyses. We briefly review it in this section.

Consider strain data $d(t)$, consisting of a modelled signal $h(t;\theta)$,
noise $n(t)$, and a deviation from the model $\delta(t; \theta_\text{True}, \beta)$
 \begin{align}
	\label{eq:strain_deviation}
	d(t)
	&=
	h(t;\theta)
	+ 
	n(t)
	+
	\delta(t; \theta_\text{True} , \beta),
\end{align}
where the source parameters $\theta$ used to generate the model may be
different from the true source parameters $\theta_\text{True}$.
We allow $\delta$ to depend on both the source parameters $\theta_\text{True}
$, as well as a new parameter $\beta$. The latter is zero when there is no
deviation.
Crucially, $\beta$ may depend on  $\theta_\text{True} $. This is the case, for example, in
\glspl{eft} of gravity, where deviations are introduced by high-order operators
that will scale as the inverse of the black hole mass (e.g.~\cite{maselliBlackHoleSpectroscopy2024,Bernard:2025dyh}), or when investigating
Lorentz violations that cause dispersion, where the deviation will change
according to redshift (e.g.~\cite{Yagi:2013ava}).

Given the residual $r^{X}(t) := d(t) - h(t; \theta)$ in detectors $X=I,J$, the \gls{crisp} is given by
\begin{align}
	\label{eq:cross_correlation_definition}
	D^{IJ}(t)
	&=
	\int^{t + \frac{\tau(t)}{2}}_{t - \frac{\tau(t)}{2}}
	r^{I}(t)
	r^{J}(t)
	dt. 
\end{align}
The binning width $\tau(t)$ is a new parameter of the analysis.
It can be tuned to balance variance reduction from
averaging over uncorrelated noise, and bias from washing out timescales smaller
than  $\tau$, as explained in~\cite{dideronNewFrameworkStudy2023}.

An important quantity to characterise is the variance $\varOfD$ of $D^{IJ}(t)$,
which can be computed from the zero-lag auto-correlation function of the noise
in each detector:
\begin{align}
	\varOfD (t)
	:=
	\langle D (t) D (t) \rangle
	&=
	\frac{1}{n_\text{eff}}
	\left<
	n^{I}(t)^{2}
	\right>
	\left<
	n^{I}(t)^{2}
	\right>,
\end{align}
where $n_\text{eff}$ is the effective number of samples in a bin of size
$\tau$, and accounts for the auto-correlation of the noise within this bin,
see~\cite{brockwellTimeSeriesTheory2013}. The angle brackets denote the expectation value over noise realizations. 
In order to recover the parameter $\beta$, we can use power templates
$Z(t;\beta)$ that model the \gls{crisp} to construct a likelihood:
\begin{align}
	\ln
	\Lk \left( D^{IJ} | \theta, \beta \right)
	&\ni
	- \frac{1}{2}
	\int
	dt
	\frac{D^{IJ}(t) Z(t;\beta)}{\langle K_{D} (t) K_{D} (t) \rangle},
	\label{eq:crisp_likelihood}
\end{align}
where the form of the likelihood follows from assuming that, in the absence of
a signal, the correlation between $D^{IJ}(t_1)$ and $D^{IJ}(t_2)$ is negligible for
$t_1 \neq t_2$. The optimal estimator for the assumed noise statistics (Wiener filter)
is also used, as discussed in Sec.~\ref{app:unmodelled_inference}.

The \gls{crisp} is essentially a compressed signal, and
different strain signals may lead to the same \gls{crisp}. 
Similarly, each power template $Z(t;\beta)$ corresponds to multiple possible
strain deviations, and represent general features expected from a family of
physically motivated deviations.

Power templates are less informative than strain models (i.e.
$\delta$).
At best, the likelihood on the \gls{crisp} given a power template
will be as informative as that of strain data given a precise model of
the strain deviation.~\footnote{This is just an example of the data processing
inequality~\cite{shannonMathematicalTheoryCommunication1948}.
Intuitively, the averaging operation means that each bin is less
sensitive to $\beta$, and thus reduces the Fisher information and the
KL divergence between values of $\beta$.}
The advantage of power templates and the \gls{crisp} is that they do not
require modelling of the strain deviation\footnote{An interesting possiblity for future work is to explore
the possibility of phase retrieval as discussed
for intensity data in the electromagnetic 
sector, e.g.~\cite{Fienup:82,Dong_2023,Dalal:2024aaj}.}.
This is particularly useful when we are interested in types of deviations that
vary across a population of events, as it allows us to constrain population
effects without relying on a precise strain model.
We describe how such effects can be measured in the next section.

\subsection{Bayesian Framework}%
\label{sub:Bayesian Framework}

We denote the parameters describing the variation of the deviation across the 
population as $\Delta_{m}$, such that there exists a distribution 
$P \left( \beta | \Delta_{m}, \theta \right)$. Given a set of data $\{ d \}_{i}$ corresponding to $N_\text{event}$ events, the
posterior on the parameters $\Delta_{m}$ is \cite{dideronNewFrameworkStudy2023}
\begin{align}
&\Po
\left( \Delta_{m} | \{ d \}_{i}  \right) 
\\
\propto
&\prior \left( \Delta_{m} \right)
\prod^{N_\text{event} }_{i=1}
\prod^{N_\text{det}}_{I\neq J}
\frac{
\Lk \left( D^{IJ}_{i} |  \Delta_{m} \right)
}{
\xi \left( \Delta_{m} \right)
}
\\
\propto
&\prior \left( \Delta_{m} \right)
\prod^{N_\text{event} }_{i=1}
\prod^{N_\text{det}}_{I\neq J}
\frac{
\int d \theta 
\Lk \left( D^{IJ}_{i}| \theta, \Delta_{m} \right)
\prior \left( \theta \right)
}{
\xi \left( \Delta_{m} \right)
}
\label{eq:full_hierarchical_posterior}
\\
\propto
&\prior \left( \Delta_{m} \right)
\prod^{N_\text{event} }_{i=1}
\prod^{N_\text{det}}_{I\neq J}
\frac{
\int d \theta d \beta
\Lk \left( D^{IJ}_{i}| \theta, \beta \right)
\prior \left( \beta| \theta, \Delta_{m} \right)
\prior \left( \theta \right)
}{
\xi \left( \Delta_{m} \right)
},
\end{align}
where the \gls{crisp} $D^{IJ}_{i}$ is computed for each possible pair $I$ 
and $J$ among the $N_\text{det}$ detectors, for each event $i$.
The prior on the source parameters $\prior \left( \theta \right)$ is obtained
from previous parameter estimation.
Finally, $\xi$ is the proportion of events that are detected given the parameters
$\Delta_{m}$, and accounts for selection effects:
\begin{align}
	\xi \left( \Delta_{M} \right)
	&=
	\int d \theta
	d \beta
	p
	\left(
		\text{detected}
		|
		\theta,
		\beta
	\right)
	\prior
	\left( \beta | \theta, \Delta_{M} \right)
	\prior
	\left( \theta \right).
\end{align}

The \gls{crisp} likelihood $\Lk \left( \{d^{I}, d^{J}\}_{i} | \theta,
\Delta_{m} \right)$ is more agnostic than a strain likelihood.
The inference can further be freed of assumptions 
by using the Wiener filter of the deviation in place of the deviation parameter $\beta$.
We call such a search an \emph{unmodeled search}, and describe it in the next
section.

\subsection{Unmodeled inference}%
\label{app:unmodelled_inference}

When evaluating the  likelihood~\eqref{eq:crisp_likelihood} it is not necessary to assume a
parametrised form for the deviation at the strain level, which provides the
parameter $\beta$ that measures the strength of the deviation from the model.
Instead, we can measure the mean power spectrum over bins.
For example, in~\cite{dideronNewFrameworkStudy2023}, we assume that the \gls{crisp}
is ``chirp-like'' and increases close to merger.
In such cases, $Z(t)$ depends on the same parameters $\theta$ as the model.
To measure the strength of the deviation, we cannot use $\beta$,
and instead measure the Wiener filter statistic $\alpha$:
\begin{align}
	\alpha
	&:=
	\frac
	{\int dt D(t) Z(t; \theta)}
	{\sqrt{\int dt \var D(t) Z(t; \theta)^2}}
	,
	\label{eq:wiener_statistics}
\end{align}
which is a measure of the \gls{snr}.

The likelihood $\Lk \left( \{d^{I}, d^{J}\}_{i} | \theta, \Delta_m \right)$ is
then marginalized over $\alpha$ rather than $\beta$. The integration is
trivial as $\Lk \left( \{d^{I}, d^{J}\}_{i} | \theta, \Delta_m \right)$ is a
normalized Gaussian with mean $\alpha$ by definition of the \gls{snr}.

An important point is that the \gls{snr} of an agnostic template will not
necessarily scale with the power of the deviation. That is, $\alpha$ may not be
proportional to $\beta$.
This is most easily seen in the extremum case: if $Z(t)$ and the true deviation
do not overlap in time, then $\alpha$ will always be zero regardless of the
value of $\beta$.

An agnostic template $Z(t)$ will only be sensitive to deviations with which it
overlaps, and in the limit of a small deviation.
To explain the last point, assume that we can decompose the cross-correlated residuals as a product of 
a time and a model-parameters dependent function $F(t)$, and a parameter $\beta$
that measures the strength of the deviation from the model: $D(t) = \beta^{2}(\theta) F(t)$.
Then, 
\begin{align}
	\alpha
	&=
	\beta^{2}(\theta)
	\frac
	{\int dt  F(t) Z(t; \theta)}
	{\sqrt{\int dt \var D (t) Z(t; \theta)^2}},
\end{align}
and we see that whether $\alpha$ scales with $\beta$ depends on the
noise-weighted overlap of the template $Z(t; \theta)$ with the function $F(t)$.

\section{Application of \score on \gls{gwtc3}}%
\label{sec:Application on GWTC-3}

\begin{table}[htbp]
\centering

\begin{NiceTabular}{lrrrr}[code-before=\rowcolors{2}{gray!25}{white}]

\toprule
Event & $m_1^{\rm src}\,[M_\odot]$ & $m_2^{\rm src}\,[M_\odot]$ & SNR & $\alpha$ \\
\midrule
GW150914 & $35.6^{+4.7}_{-3.1}$ & $30.6^{+3.0}_{-4.4}$ & $26.0^{+0.1}_{-0.2}$ & $0.55$ \rs
GW151226 & $13.7^{+8.8}_{-3.2}$ & $7.7^{+2.2}_{-2.5}$ & $13.1$ & $1.10$ \rs
GW170104 & $30.8^{+7.3}_{-5.6}$ & $20.0^{+4.9}_{-4.6}$ & $13.8^{+0.2}_{-0.3}$ & $-0.97$ \rs
GW170608 & $11.0^{+5.5}_{-1.7}$ & $7.6^{+1.4}_{-2.2}$ & $15.4$ & $-0.49$ \rs
GW170809 & $35.0^{+8.3}_{-5.9}$ & $23.8^{+5.1}_{-5.2}$ & $12.8^{+0.2}_{-0.3}$ & $0.98$ \rs
GW170814 & $30.6^{+5.6}_{-3.0}$ & $25.2^{+2.8}_{-4.0}$ & $17.7^{+0.2}_{-0.3}$ & $0.83$ \rs
GW170823 & $39.5^{+11.2}_{-6.7}$ & $29.0^{+6.7}_{-7.8}$ & $12.2^{+0.2}_{-0.3}$ & $1.38$ \rs
GW190408\_181802 & $24.8^{+5.4}_{-3.5}$ & $18.5^{+3.3}_{-4.0}$ & $14.6^{+0.2}_{-0.3}$ & $-1.22$ \rs
GW190412\_053044 & $27.7^{+6.0}_{-6.0}$ & $9.0^{+2.0}_{-1.4}$ & $19.8^{+0.2}_{-0.3}$ & $-1.15$ \rs
GW190503\_185404 & $41.3^{+10.3}_{-7.7}$ & $28.3^{+7.5}_{-9.2}$ & $12.2^{+0.2}_{-0.4}$ & $-0.69$ \rs
GW190512\_180714 & $23.2^{+5.6}_{-5.6}$ & $12.5^{+3.5}_{-2.6}$ & $12.7^{+0.3}_{-0.4}$ & $0.09$ \rs
GW190513\_205428 & $36.0^{+10.6}_{-9.7}$ & $18.3^{+7.4}_{-4.7}$ & $12.5^{+0.3}_{-0.4}$ & $1.33$ \rs
GW190519\_153544 & $65.1^{+10.8}_{-11.0}$ & $40.8^{+11.5}_{-12.7}$ & $15.9^{+0.2}_{-0.3}$ & $0.66$ \rs
GW190521\_030229 & $98.4^{+33.6}_{-21.7}$ & $57.2^{+27.1}_{-30.1}$ & $14.3^{+0.5}_{-0.4}$ & $-0.66$ \rs
GW190521\_074359 & $43.4^{+5.8}_{-5.5}$ & $33.4^{+5.2}_{-6.8}$ & $25.9^{+0.1}_{-0.2}$ & $0.66$ \rs
GW190602\_175927 & $71.8^{+18.1}_{-14.6}$ & $44.8^{+15.5}_{-19.6}$ & $13.2^{+0.2}_{-0.3}$ & $0.14$ \rs
GW190706\_222641 & $74.0^{+20.1}_{-16.9}$ & $39.4^{+18.4}_{-15.4}$ & $13.4^{+0.2}_{-0.4}$ & $0.25$ \rs
GW190707\_093326 & $12.1^{+2.6}_{-2.0}$ & $7.9^{+1.6}_{-1.3}$ & $13.1^{+0.2}_{-0.4}$ & $-1.53$ \rs
GW190728\_064510 & $12.5^{+6.9}_{-2.3}$ & $8.0^{+1.7}_{-2.6}$ & $13.1^{+0.3}_{-0.4}$ & $1.44$ \rs
GW190814\_211039 & $23.3^{+1.4}_{-1.4}$ & $2.6^{+0.1}_{-0.1}$ & $25.3^{+0.1}_{-0.2}$ & $1.23$ \rs
GW190828\_063405 & $31.9^{+5.4}_{-4.1}$ & $25.8^{+4.9}_{-5.3}$ & $16.5^{+0.2}_{-0.3}$ & $1.37$ \rs
\textbf{GW190915\_235702} & $32.6^{+8.8}_{-4.9}$ & $24.5^{+4.9}_{-5.8}$ & $13.1^{+0.2}_{-0.3}$ & $2.47$ \rs
GW191109\_010717 & $65.0^{+11.0}_{-11.0}$ & $47.0^{+15.0}_{-13.0}$ & $17.3^{+0.5}_{-0.5}$ & $0.01$ \rs
GW191129\_134029 & $10.7^{+4.1}_{-2.1}$ & $6.7^{+1.5}_{-1.7}$ & $13.1^{+0.2}_{-0.3}$ & $0.59$ \rs
\textbf{GW191204\_171526} & $11.7^{+3.3}_{-1.7}$ & $8.4^{+1.3}_{-1.7}$ & $17.4^{+0.2}_{-0.3}$ & $2.27$ \rs
GW191222\_033537 & $45.1^{+10.9}_{-8.0}$ & $34.7^{+9.3}_{-10.5}$ & $12.5^{+0.2}_{-0.3}$ & $-1.28$ \rs
\textbf{GW200129\_065458} & $34.5^{+9.9}_{-3.1}$ & $29.0^{+3.3}_{-9.3}$ & $26.8^{+0.2}_{-0.2}$ & $-3.22$ \rs
GW200224\_222234 & $40.0^{+6.7}_{-4.5}$ & $32.7^{+4.8}_{-7.2}$ & $20.0^{+0.2}_{-0.2}$ & $1.49$ \rs
GW200225\_060421 & $19.3^{+5.0}_{-3.0}$ & $14.0^{+2.8}_{-3.5}$ & $12.5^{+0.3}_{-0.4}$ & $-1.41$ \rs
\textbf{GW200311\_115853} & $34.2^{+6.4}_{-3.8}$ & $27.7^{+4.1}_{-5.9}$ & $17.8^{+0.2}_{-0.2}$ & $-2.53$ \rs

\bottomrule

\end{NiceTabular}

\caption{Selected events from
	GWTC-3~\cite{ligoscientificcollaborationGWTC3CompactBinary2023} with inferred \gls{mle} of the
	component masses (with 90\% \gls{ci}), network \gls{snr} and Wiener
	filter statistic $\alpha$. Events in bold are those with $\alpha$ above
2, and are further discussed in the text.}
\label{tab:gwtc-summary}

\end{table}

In this section, we apply the \score framework to \gls{gwtc3}~\cite{ligoscientificcollaborationGWTC3CompactBinary2023}\footnote{We
chose to apply the method to the latest data release for which Testing-\gls{gr}
analyzes had been published \cite{LIGOScientific:2021sio} before the completion
of the work. Testing-\gls{gr} analyzes have now been applied to
the Fourth Gravitational-Wave Transient Catalog (GWTC-4)~\cite{abacGWTC40IntroductionVersion2025,theligoscientificcollaborationGWTC40UpdatingGravitationalWave2025,theligoscientificcollaborationGWTC40TestsGeneral2026,theligoscientificcollaborationGWTC40TestsGeneral2026a,theligoscientificcollaborationGWTC40TestsGeneral2026b}.}
We envision two different
scenarios where a potential deviation can have a mass dependency in the population-based analysis described in section~\ref{sec:Method}. The first, and the one most relevant for this work, is an unmodelled case. Then, we also briefly
examine a case where a specific model is assumed.

We chose events with network match filtering \gls{snr} above 12, and for which
strain data is available in both LIGO Hanford and Livingston detectors.
The resulting list of 30 events is presented in Table~\ref{sec:Application on GWTC-3}.

\begin{figure*}
	\begin{minipage}{0.69\textwidth}
		\begin{center}
			\includegraphics[width=\columnwidth]{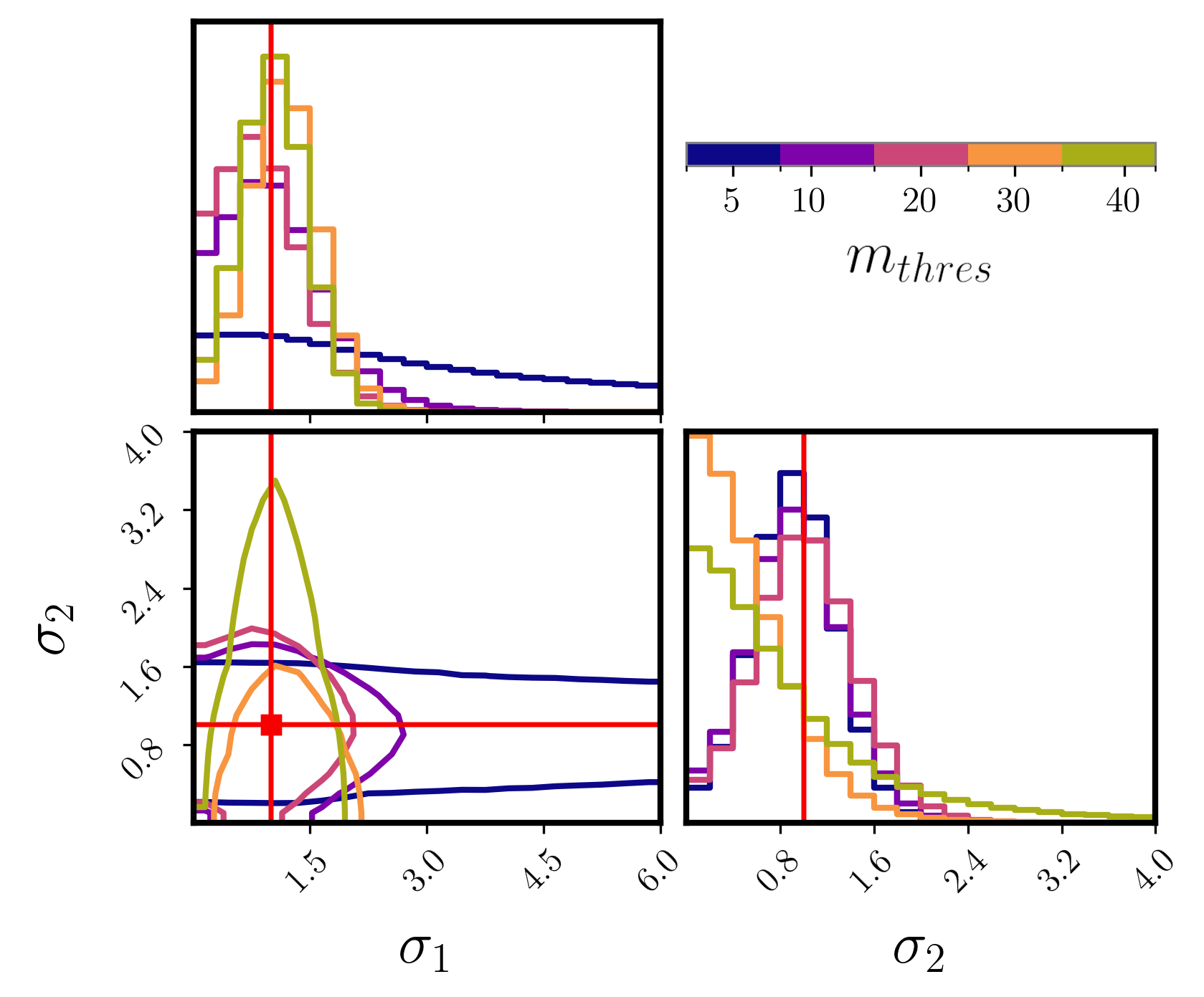}
		\end{center}	
	\end{minipage}
	\\
	\begin{minipage}{0.49\textwidth}
		\begin{center}
			\includegraphics[width=\columnwidth]{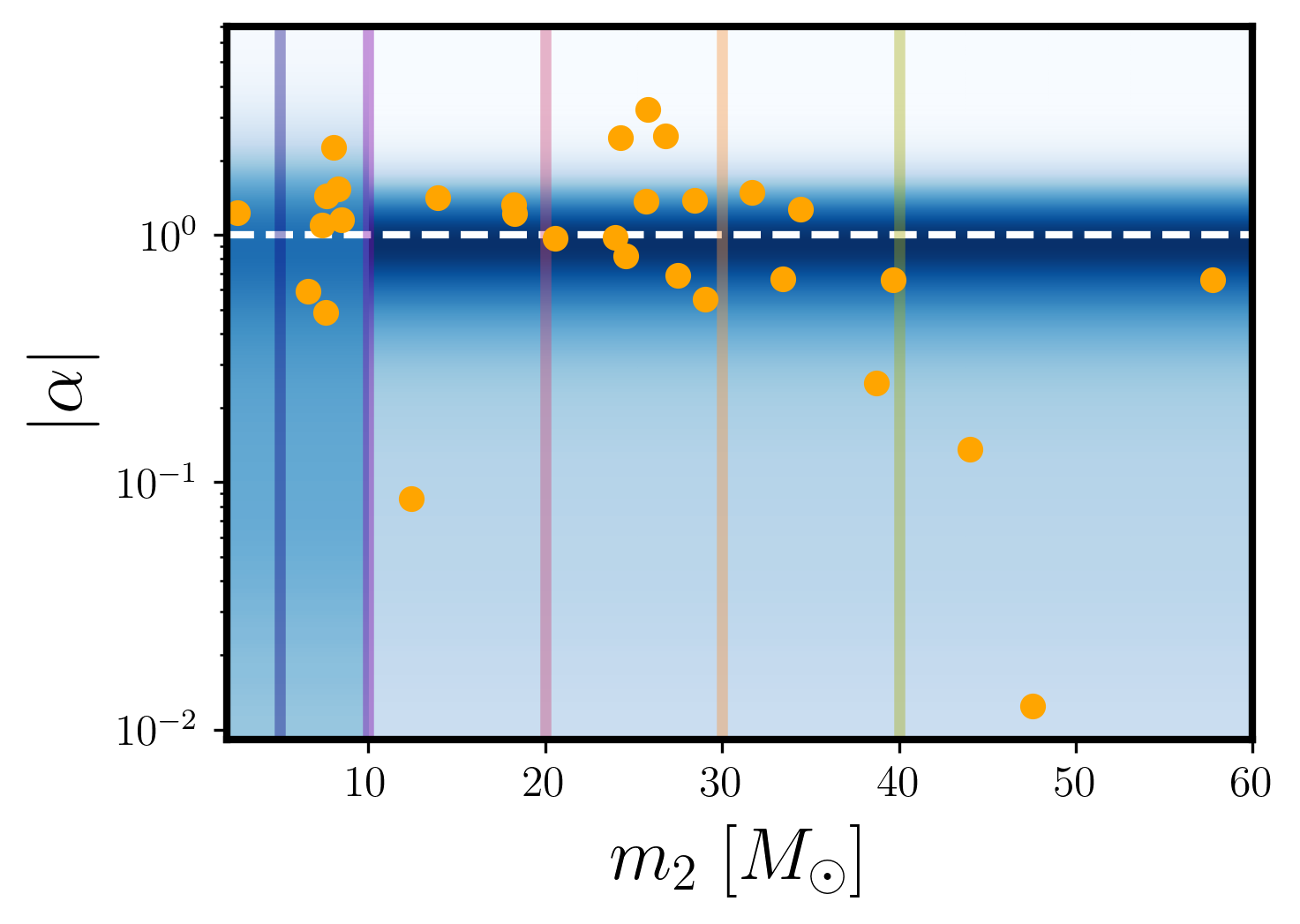}
		\end{center}	
	\end{minipage}
	\begin{minipage}{0.49\textwidth}
		\begin{center}
	\includegraphics[width=\columnwidth]{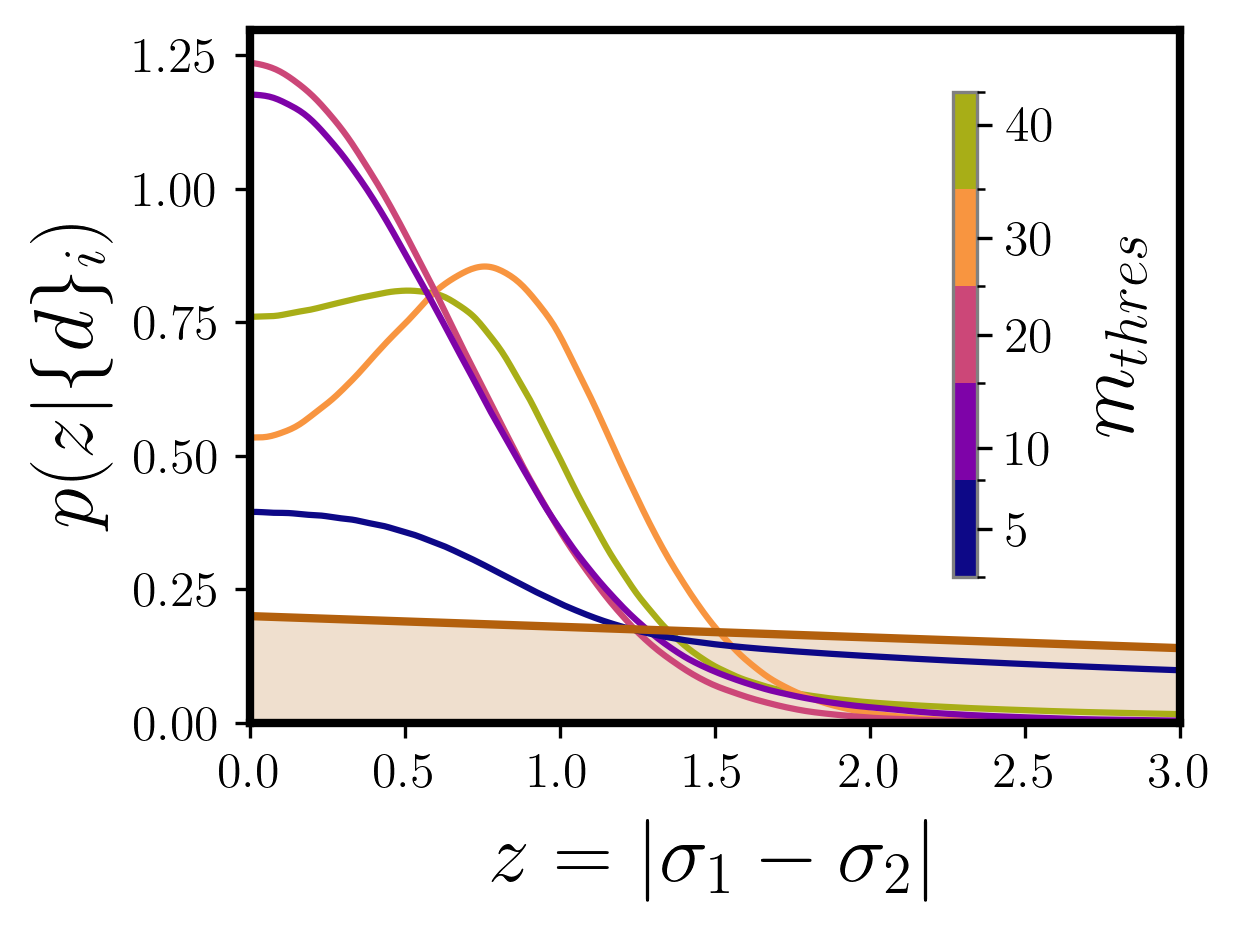}
		\end{center}	
	\end{minipage}
	\caption{Result of the unmodeled search. The top panel shows the joint
		posterior $\Po \left( \sigma_{1}, \sigma_{2} | \{d\}_i \right)$ 
		on the standard deviations of the low and high mass
		populations for different definitions of the mass threshold on
		$m_{2}$ separating the two populations. The prior flat, assuming both $\sigma_1$ and $\sigma_2$ are between 0 and 10.
		For all values of the threshold, the noise model 
		($\sigma_1=\sigma_2=1$) is recovered within the 90\% \gls{ci}.
		The bottom left panel shows the distribution of the Wiener
		filter statistic $\alpha$ as a function of the \gls{mle} value of the
		smaller component mass $m_{2, \text{MLE}}$. 
		The background shows the marginalized posteriors on $\sigma_{1}$
		and $\sigma_{2}$ for the case $m_\text{threshold}=10
		M_{\odot}$.
		The bottom right panel shows that the posterior
		on the difference $z:=|\sigma_1 - \sigma_2|$ has strong support
		for $z=0$, i.e. the two populations having the same variance. The brown shaded area shows the prior on $z$, given the prios on  $\sigma_1$ and $\sigma_2$.
		Higher mass thresholds have support further from
		$z=0$---this is due to the 3 low $\alpha$ events 
		GW191109\_010717, GW190602\_175927 and GW190706\_222641.
	}
	\label{fig:mass_gap_results}
\end{figure*}

\subsection{Unmodelled search}%
\label{sec:Unmodelled search}

Let us examine a scenario where the underlying strain deviation is
not known or assumed, so we use an unmodelled search.
We make the minimalist assumption that such deviation is reflected on a mass population effect, and use a simple model that  assumes two populations split by a
mass threshold.
This framework can be used, for example, to investigate whether all compact
objects in observed events are really black holes. Such a scenario could arise within General Relativity
with the potential existence of boson stars. Regardless of the specific details of the different models proposed
for them, all have a maximum mass above which only black hole solution exists (see, e.g.~\cite{Liebling:2012fv}). The exact nature of the
object, which would influence the strain deviation, does not need to be known,
and these objects may be suspected to populate different mass ranges (for a specific, modelled search 
for binary boson stars see~\cite{Evstafyeva:2024qvp}).

We posit a simple population model to capture the scenario we described above. We assume the presence of two population of compact objects, each resulting in a different distribution of deviations from the model. They are characterized by a mass threshold $\mthreshold$ separating the two populations. We set the threshold in terms of the smaller component mass $m_{2}$, as this is the mass that sets the highest curvature scale of the system. When we measure a deviation in the form of the Wiener filter statistic $\alpha$, the measurement will be drawn from one of these two distributions, depending on which type of compact object is present in the binary. By definition of the Wiener filter statistics, both of these distributions are Gaussian, but their mean and standard deviation may differ. We fix the mean to be zero. This reduces computational complexity, and is justified by the fact that no event yet has shown a significant deviation from the model. Each population is then described by its standard deviation: $\sigma_{1}$ when $m_{2} < \mthreshold$ and $\sigma_{2}$ when $m_{2} \geq \mthreshold$. 

The prior in Eq. ~\eqref{eq:full_hierarchical_posterior} is:
\begin{align}
	\label{eq:}
	\prior \left( \alpha| \theta, \Delta_{M} \right)
	&=
	\frac{1}{\sqrt{2 \pi \sigma^{2}}}
	\exp
	\left(
		- \frac{1}{2}
		\frac
		{
			\alpha^{2}
		}
		{\sigma^{2}}
	\right)
	,\\
	\sigma
	&=
	\begin{cases}
		\sigma_{1}, & m_{2} <
		m_\text{threshold},\\
		\sigma_{2}, & m_{2} \geq
		m_\text{threshold},
	\end{cases}
\end{align}
while we set the hyperprior on $\sigma_{1}$ and $\sigma_{2}$ to be flat in
the range $[0, 10]$.
The final piece of the model is the selection effect term $\xi \left(
\Delta_{m} \right)$. Since we focus on high \gls{snr} events and assume that the deviations are too small to affect the detection statistics of the events, $\xi$ does not depend on $\Delta_M$. Its computation may become relevant when large deviations are possible, for example when searching for completely unmodeled signals.

For computational simplicity, we assume that the MLE of the source parameters
is the true source parameters, and set $\prior \left( \theta \right) = \delta
\left( \theta_\text{MLE}  \right)$.
The simplicity of this model allows us to compute the integral in Eq. ~\eqref{eq:full_hierarchical_posterior} analytically, giving the following
event likelihood:
\begin{align}
	&\Lk
	\left( 
		D^{IJ}_{i}| \Delta_{m}=\{\sigma_{1}, \sigma_{2}\}
	\right)
	\\
	=&
	\frac{1}{\sqrt{2 \pi \left( \sigma^{2} + 1 \right)}}
	\exp	
	\left( 
		- \frac{1}{2}
		\frac
		{
			\alpha_{i}^{2}
			\left(
				D^{IJ}_{i}
			\right)
		}
		{\sigma^{2} + 1}
	\right), \label{eq:unmodeled_pop_likelihood}
	\\
	\sigma
	&=
	\begin{cases}
		\sigma_{1}, & m_{2, \text{MLE}} <
		m_\text{threshold},\\
		\sigma_{2}, & m_{2, \text{MLE}} \geq
		m_\text{threshold}.
	\end{cases}
\end{align}
The right-hand-side is obtained from the data $d(t)$ and strain model
$h(t; \theta_\text{MLE})$ for an event $i$
by computing the \gls{crisp} as defined 
in Eq. ~\eqref{eq:cross_correlation_definition}, and then evaluating the
Wiener filter statistic $\alpha_{i}$ for the event using
Eq. ~\eqref{eq:wiener_statistics}.
We do not make any assumption about how the deviation power is distributed in time, and filter for any excess over the signal by making $Z(t; \theta)$ constant in $t$.

The resulting posteriors on $\sigma_{1}$ and $\sigma_{2}$ are shown at the top of 
Fig. ~\ref{fig:mass_gap_results}, while bottom left of the figure
shows the distribution of $\alpha$ as a function of the \gls{mle} value of the
smaller component mass. The marginalized posteriors on $\sigma_{1}$ and
$\sigma_{2}$ (the probability distribution density for the standard deviation
of $\alpha$) are plotted in the background for $\mthreshold = 10 M_{\odot}$
The important observation is that the noise model ($\sigma_1=\sigma_2=1$) is
well within the 90\% credible interval for all mass thresholds considered.

To compare the two populations, the interesting measure is the posterior on the difference $z:=|\sigma_1 - \sigma_2|$ between their
standard deviations, which we plot in the bottom right of
Fig. ~\ref{fig:mass_gap_results}, along with the prior (which follows from the
hyperprior on $\sigma_1$ and $\sigma_2$).
The case of the populations being equal ($z=0$) has strong support for all
threshold considered, with Savage-Dickey density ratios ranging from 2 to 6 in favor
of the equal-population hypothesis.  

We note that the two population hypothesis is less confidently disfavored by  extreme values of the mass thresholds.
For $\mthreshold=5 M_{\odot}$, this is due to a low number of events below $5 M_{\odot}$. 
In effect, the low number of events causes the uncertainty on $\sigma_1$ to be
too large to make any confident statement about the difference between the two
populations. This is worsened by our \gls{snr} selection criteria, as low mass
events are generally of lower \gls{snr} and are therefore suppressed.
For $\mthreshold\geq 30 M_{\odot}$, the lower Savage-Dickey density ratio is due to
three low $\alpha$ events (GW191109\_010717, GW190602\_175927, and
GW190706\_222641) dominating the high mass range. 
We note that these events are not statistical outliers. As a figure of
comparison, when 30 draws are made from a standard normal distribution, the
the probability that 3 values in a row have absolute value below 0.2 is about
8\%. 

\begin{figure*}
	\includegraphics[width=\textwidth]{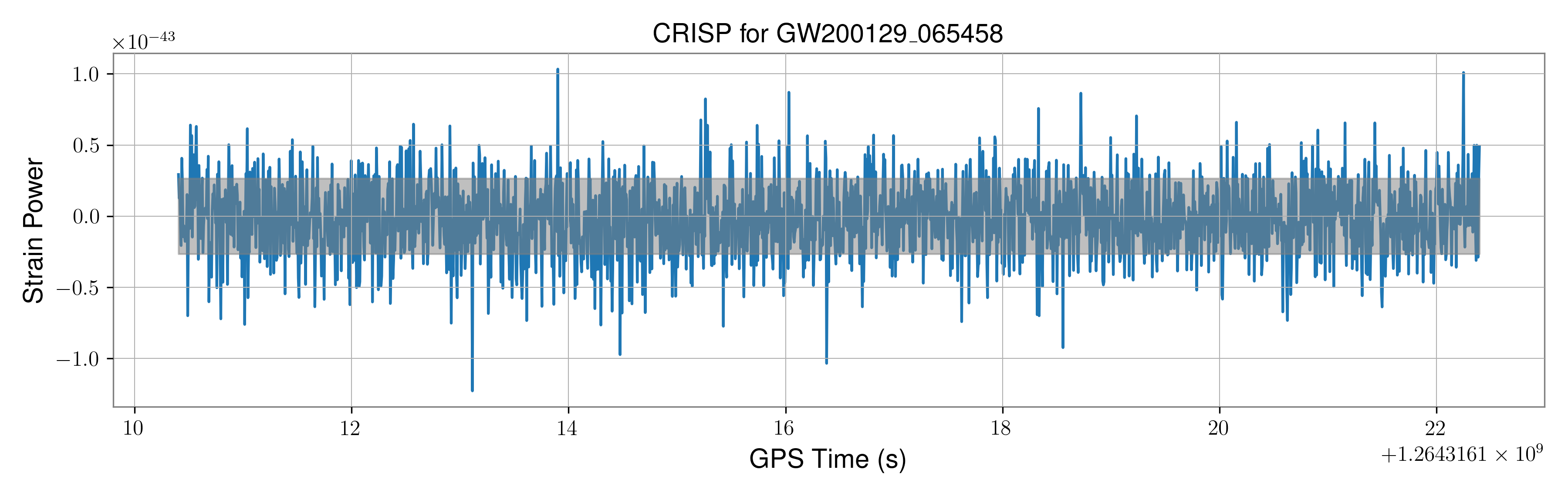}
	\caption{
		\gls{crisp} for GW200129\_065458, the event with the highest
		measured Wiener filter statistic $\alpha$.
	}
	\label{fig:gw200129_065458_crisp}
\end{figure*}

We also note the highest three $\alpha$ events: GW190915\_235702, 
GW191204\_171526, and GW200129\_065458. Although these events have $\alpha$
above 2 standard deviations, we did not observe any visible systematic
deviation in the \gls{crisp} (plotted for GW200129\_065458 in
Fig. ~\ref{fig:gw200129_065458_crisp}). It is likely that these high $\alpha$
values are due to statistical fluctuations in the noise.

\begin{figure}
	\includegraphics[width=\columnwidth]{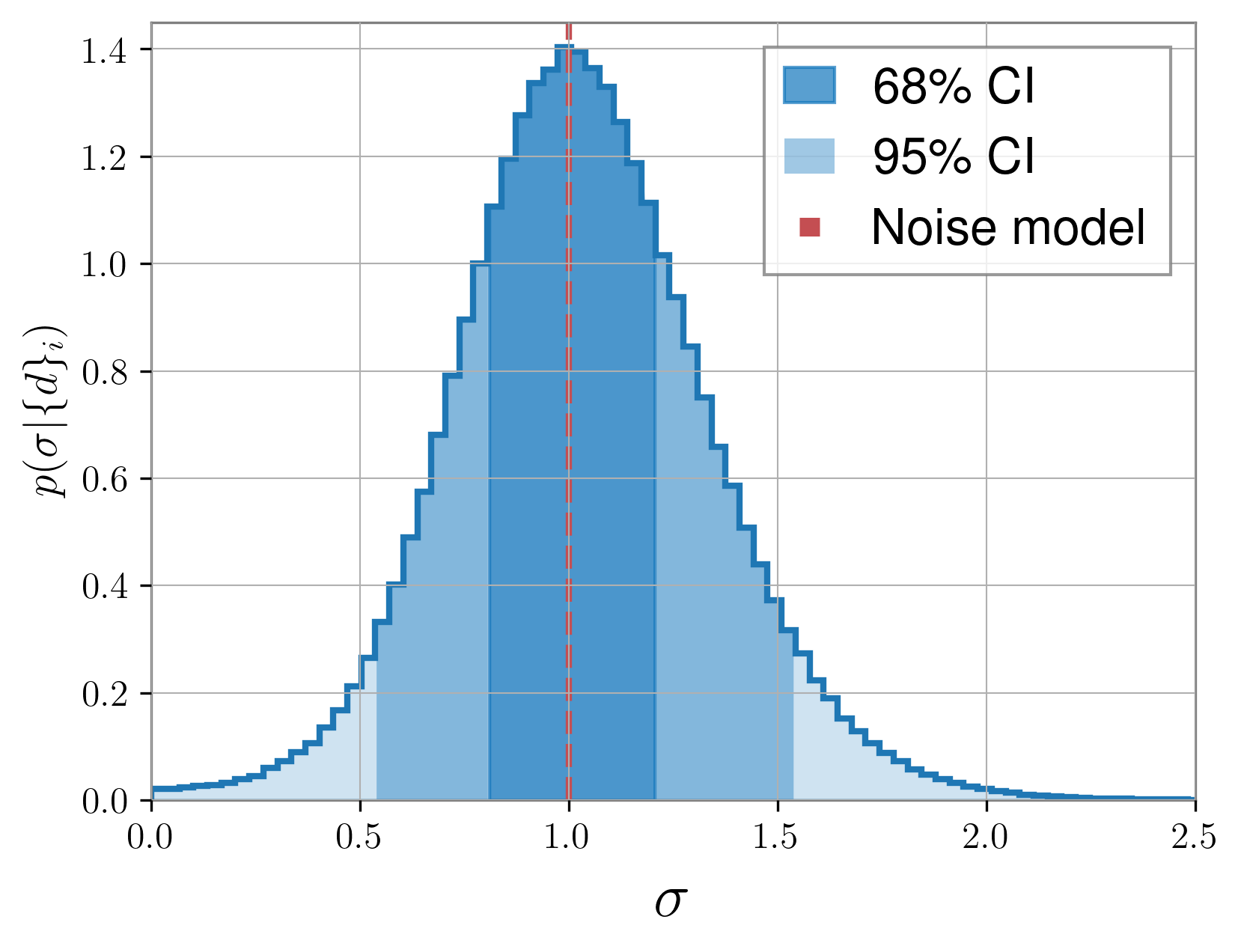}
	\caption{
		Unmodeled search with a single population model,
		where each event's deviation parameter $\alpha$ is
		assumed to be drawn from a Gaussian distribution with mean
		zero and standard deviation $\sigma$. The posterior peaks
		around the value expected in the presence of noise alone,
		$\sigma=1$. We also show the Highest Posterior Density
		intervals containing 68\% and 95\% of the posterior
		probability.
	}
	\label{fig:full_pop}
\end{figure}

Finally, for completeness, we check whether the set of events as a whole
indicates deviations from the noise model.
Fig. ~\ref{fig:full_pop} shows the posterior on the standard
deviation $\sigma$ assuming that all events belong to a single Gaussian
distribution with mean zero. The distribution peaks at the noise model value of
1. The prior is the same: a flat distribution in the range $\sigma=[0,10]$. Thus, using the
Savage-Dickey density, the Bayes factor in favor of a deviation is $\bdev=0.07$.
We therefore find that the population as a whole is consistent with the underlying noise
property and there is no statistically significant deviation indicating the presence of any mass-scale in the compact objects present in GWTC-3.

\subsection{Modelled search for EFT effects}%
\label{sec:Modelled search for EFT effects}

Here we focus on a suspected physical effect with known
strain behaviour.
As described in previous
studies~\cite{carulloEnhancingModifiedGravity2021a,maselliBlackHoleSpectroscopy2024,dideronNewFrameworkStudy2023,Bernard:2025dyh}, higher order curvature corrections to \gls{gr}
will have an effect on the \gls{gw} emission from compact binary mergers, and this effect 
scales with the mass scale of the binary.
In particular, black-holes in \glspl{eft} have non-zero tidal
deformability (e.g.~\cite{cardosoBlackHolesEffective2018}).
In non-spinning binaries, described by \gls{eft}-motivated beyond \gls{gr} theories without the addition of extra fields, such deformability mediates the leading order modification of inspiral waveforms~\cite{Bernard:2025dyh}. The impact is larger for lighter masses.
We can therefore constrain these theories by searching for tidal effects in binary black hole mergers (as opposed to neurtron star mergers), and scrutinizing any such effect for a scaling with the mass scale of the binary.

Based on these observations, we use the setup presented
in~\cite{dideronDetectingUnmodeledSourcedependent2025}, and search for 
an effective tidal-deformability that scales as a power law of the lower 
component mass of the binary.
As a quick summary of this setup, we looked for tidal effects on the strain parametrized by $\beta$ that scaled with the lower component mass, $m_2$, as a power law: $\beta = \beta_0 m_2^D$. While the ``intercept" of the power law, $\beta_0$, controls the size of the effect, the physically significant parameter is the power index $D$, which can be directly used to put contraints on the \gls{eft} responsable for the effect. 

We construct power templates $Z(t; \beta=\effL)$ which depend on
the effective tidal deformability $\effL$ of the binary.
We used the \texttt{IMRPhenomXP\_NRTidalv2} from 
\cite{dietrichImprovingNRTidalModel2019,
colleoniIMRPhenomXP_NRTidalv2ImprovedFrequencydomain2023} to obtain strain
templates, from which the \gls{crisp} templates are constructed by subtracting
the waveform with $\effL=0$ and computing
Eq.~\eqref{eq:cross_correlation_definition}.
The prior on $\effL$ for each event is deterministically set by the power law
scale and index: $\prior \left( \beta | \theta, \beta_{0}, D \right) = \delta
\left( \beta - \beta_{0} m_{2}^{D} \right)$.

We found that the posterior is uninformative on $\effL$ between $0$ and at least
$10^{5}$ for all events. This upper limit is set by the prior, which itself we
set based on existing bounds on the tidal deformability of $> 5 M_{\odot}$ event components 
previously found by template searches. For example,
\cite{chiaPursuitLoveNumbers2024,Andres-Carcasona:2025bni} constrained the effective tidal
deformability $\tilde{\Lambda}$ to below $10^{3}$ for events with chirp mass
ranging up to $10 M_{\odot}$. 
With current detector sensitivity, it is not possible to use the \gls{crisp} to constrain tidal deformability. As we have shown in~\cite{dideronDetectingUnmodeledSourcedependent2025}, next generation detectors, however, will be able to put bounds on specific theories using \score \cite{dideronDetectingUnmodeledSourcedependent2025}.
An approach that may allow constraints with current sensitivities would be to directly model the scaling of excess power (instead of relying on strain model).

\section{Conclusion}%
\label{sec:Conclusion}

In this work, we have applied the \score framework to search for mass-dependent
deviations from waveform models in the \gls{gwtc3} catalog of events, finding
consistency with expected noise.
The \score search we performed is agnostic to the specific form of the deviation in the strain, and instead projects a family of strain deviations onto cross-correlated power space.
This is an important aspect of the method, as it allows searches for population effects that are not dependent on specific strain models, sparing modeling costs and trading specificity for generality.
Instead, we rely on physical models, on which we can put constraints.

As an example of a physical argument, several scenarios predict deviations that are dependent on mass scale (e.g. gravitational theories formulated through \glspl{eft} arguments and environmental effects,  see \cite{Bernard:2025dyh,Barausse:2014tra}).
We focused on two such cases. In the first, we explored a discrete difference between masses greater and smaller than given thresholds. 
Such a situation would arise, for example, with exotic compact
objects that exist only below a certain mass.
We separated events into
high and low mass bins, and tested whether deviations were distributed
differently in the two bins by comparing their standard deviations while the
mean was fixed.
For generality, we quantified deviations with the total cross-correlated residual power over the entire signal duration. 
This was done to minimize assumptions, but additional physical assumptions
can be made while keeping the method model agnostic, as we mention below.
We found that the two populations are consistent with each other, and with the
noise model. This was the case for multiple choices of the mass threshold
separating the two populations, and for the population as a whole. Thus, we did
not find evidence for excess cross-correlated power, nor any variation thereof with mass in the catalog.

A particular example of mass dependent deviations are gravitational theories captured by \glspl{eft}
without the addition of extra degrees of freedom (beyond the metric ones). Here deviations scale as a power-law and leading order deviations imprint as tidal effects~\cite{Bernard:2025dyh}.
To test this kind of models, we assumed the residual power arose from a
tidal-like deformation at the strain level, and computed a posterior over the index of the power-law, which leads to direct constraints on the parameters of the
theory\footnote{See~\cite{Payne:2024yhk} for an example that examines this scenario at the level of the strain.}.
We found that the current noise levels of the detectors do not allow us to constrain the
effective tidal deformability for any individual event, which prevents us from
performing population inference on this parameter. This is due to the loss of
information occasioned by the \gls{crisp}, which is necessary to perform
agnostic searches. Future detectors with lower noise curves will be able to
constrain the effective tidal deformability for individual events, enabling
population inference. 

As a final note leading to future applications,
we note that the power-law physical argument does not
demand the tidal-like form of the deviation but is used together with 
relevant information on the frequency dependency
of relevant effects, see \cite{Bernard:2025dyh,Barausse:2014tra,Cardoso:2021wlq}. The analysis could be repeated
by building a power template based on these arguments, which may allow constraints to be put on the power-law with current data.
However, even assuming a strain-level tidal model, future generation
detectors, such
as Cosmic Explorer
(CE)~\cite{reitzeCosmicExplorerContribution2019} and the Einstein
Telescope~\cite{punturoEinsteinTelescopeThirdgeneration2010}, will have
sufficient sensitivity to constrain specific theories using \score through
constraints on power-law indices, as we have shown
in~\cite{dideronDetectingUnmodeledSourcedependent2025}.

\acknowledgments
The authors would like to thank Elise S{\"a}nger for carefully reviewing the manuscript as a part of the LVK Publications \& Presentations review and providing useful comments. We also thank Reed Essick, Tamara Evstafyeva, Max Isi and Harrison Siegel for discussions.
This work was supported in part by the Natural Sciences and Engineering Research Council
(NSERC) of Canada and the Simons Foundation through
Award SFI-MPS-BH-00012593-12 (LL). LL also thanks financial support via the Carlo Fidani Rainer Weiss Chair
at Perimeter Institute and CIFAR. This research was supported in part by Perimeter Institute for
Theoretical Physics. Research at Perimeter Institute is supported in part by the Government of
Canada through the Department of Innovation, Science and Economic Development and by the
Province of Ontario through the Ministry of Colleges and Universities. This work is a part of the $\langle \texttt{data|theory}\rangle$ \texttt{Universe-Lab} which is supported by the TIFR and the Department of Atomic Energy, Government of India. This research is supported by the Prime Minister Early Career Research Award, Anusandhan National Research Foundation, Government of India. The authors would like to thank the  LIGO/Virgo scientific collaboration for providing the GW strain data. LIGO is funded by the U.S. National Science Foundation. Virgo is funded by the French Centre National de Recherche Scientifique (CNRS), the Italian Istituto Nazionale della Fisica Nucleare (INFN), and the Dutch Nikhef, with contributions by Polish and Hungarian institutes. This material is based upon work supported by NSF’s LIGO Laboratory, which is a major facility fully funded by the National Science Foundation. 
In the analysis done for this paper, we have used the
following packages: \textsc{PyCBC}~\cite{nitzGwastroPycbcV22022}, \textsc{LALSuite}~\cite{lalsuite}, 
\textsc{NumPy}~\cite{harris2020array}, \textsc{SciPy}~\cite{2020SciPy-NMeth}
and \textsc{Matplotlib}~\cite{Hunter:2007} with
\textsc{Seaborn}~\cite{Waskom2021}.

\bibliography{main.bib}

\appendix

\end{document}